\documentclass[showpacs]{revtex4}
\usepackage{epsfig}
\begin{document}

\title{\sc
The $D-{\bar{D}}$ mesons matter in Walecka's mean field theory}

\author{M.L. de Farias Freire}
\email{morgana.ligia@bol.com.br}
\affiliation{Departamento de F\'\i sica, Universidade Estadual da 
Para\'\i ba, 58.109-753 Campina Grande, PB, Brazil}

\author{R. Rodrigues da Silva}
\email{romulo@df.ufcg.edu.br}
\affiliation{Unidade Acad\^emica de F\'\i sica, Universidade Federal de Campina Grande, 58.051-970 Campina Grande, PB, Brazil}

\begin{abstract}
We study the $D-\bar{D}$ mesons matter in the framework of 
$\sigma$ and $\omega$ meson exchange model using Walecka's 
mean field theory. We choose the equal number of D and anti-D meson
then we get $\langle\omega^0\rangle=0$ and the $\langle\sigma\rangle$ field exhibits a critical temperature 
around 1.2 GeV. 
We investigate effective mass, pressure, energy density and
energy per pair.
We conclude that this matter is a gas and 
these results are not favorable for the existence 
of $D-\bar{D}$ bound state.

{\it Note: in arXiv:1211.5505 this interpretation 
has been updated where these results 
are favorable for the existence 
of $D-\bar{D}$ bound state.
}

\end{abstract}

\maketitle

\section{Introduction}

The possibility to found exotic mesons 
has been motivated many theoretical and experimental effort
to understand the structure of the new mesons. 
The new mesons are called exotic, 
because they cannot be explained in terms 
of quark-antiquark picture.
Since 2003, there are many candidates for exotic mesons
called $X$, $Y$ and $Z$, have been discovered in $B$ mesons decays \cite{x3872,y3930,z4430,y4140}.
The nature of the new mesons are completely open and there
are already many theoretical interpretations about their structure.
One of these interpretations are that their structure
is a bound state of two other mesons. 
It is analogous to a proton and a neutron 
binding together to form a deuteron. 
This idea is not new and it was studied by 
T\"ornqvist in 1991 \cite{torn1}
and he called these states as deuson.

In a second paper in 1994, 
T\"ornqvist \cite{torn2} used a meson potential model with one-pion-exchange interactions 
to study charm meson molecules with isospin I = 0:
$D^{*}\bar{D}/D\bar{D}^{*}$ and $D^{*}\bar{D}^{*}$.
Zhang and Huang \cite{zhang} calculated in QCD sum rules 
many molecular states, among then, $D\bar{D}$.
The $D\bar{D}$ state has already predicted by Gamermann {\it et al.} \cite{Gamermann} in 2007. They used the unitarization, in couple channels, of the chiral perturbation amplitudes
and this state was called X(3700) and it has $\mbox{I}$=0. 
In recent paper, Gamermann {\it et al.} \cite{Gamermann2} 
have suggested that it is possible to 
observe X(3700) in radiative decay, $\psi(3770)$ into X(3700)+$\gamma$.

In 1974, Walecka \cite{walecka} constructed an effective 
Lagrangian composed of a baryon field and two mediated
mesons $\sigma$ and $\omega$ to describe nucleon dynamics 
within highly condensed objects such a nucleus or neutron stars. 
The central idea of this model is the mean field theory (MFT), whereby we replace the meson fields operators by their expectation values.
The resulting equation of state for the system where the number of protons is equal the number of neutrons, called nuclear matter, exhibits nuclear saturation, 
where the two coupling constants in this theory 
was chosen to fit them to the binding energy and 
density of nuclear matter.
A prediction of this theory 
at zero temperature T=0
is the neutron matter is unbound.

Although this theory is trustful for large number mass,
suggest that may exist in nuclear matter clusters of nucleons
type: n-p, n-p-n, p-p-p-n, n-n-p-p, ..., on the other hand 
the in neutron matter the interaction among neutrons
are weaker than in the nuclear matter case,
so in neutron matter is more favorable n-n  states 
than states with multiple nucleons.
Experimentally states (n-p) can be
interpreted as deuterons and (n-n-p-p)
can be interpreted as alpha particle
and both particles are found in nuclear matter \cite{Blaschke:2008uf,Beyer:2000ds}.
On the other hand the state (n-n) there is no 
experimental evidence.
The other states that can be formed
within the nuclear may be interpreted as
excited states of nucleon and exotic mesons.

Another prediction of this theory at T=0 is that the nuclear matter 
is a liquid and the increasing temperature it has 
a liquid-gas phase transition \cite{walecka}.
The critical point has been measured by several experiments \cite{Viola:2003hy} and this value is T=$(7 \pm 1)$ MeV.

For the pion matter, this situation is opposite case that of nuclear matter.
Shuryak \cite{Shuryak:1990ie} predicted that this system at finite temperature is a liquid. 
Kostyuk {\it et al.} \cite{Kostyuk:2000if} 
have been proposed that the equation of state
of pion matter gives a 
phase transition gas-liquid at T\textless 136 MeV.
More recently, Anchishkin and Nazarenko \cite{Anchishkin:2006nx}
used mean field theory and predicted that the temperature of 
the phase transition gas-liquid is T=43 MeV and
the coupling constant was extracted 
for using the data of $\pi^{+}\pi^{-}$ 
invariant mass spectrum. 

In this work, we apply the Walecka's mean field theory
to study the $D-\bar{D}$ matter. 
We focus our attention on the existence of $D-\bar{D}$ 
bound state or X(3700). 
We consider if this bound state exists, then
this matter should has a phase transition gas-liquid. 

\section{Theory}

Recently, Ding \cite{ding} studied the meson-meson 
system, $Y(4260)$ and $Z^{+}_{2}(4250)$, 
by using a procedure for converting
a $T$-matrix into an effective potential, 
where the strong interactions 
are generated by 
$\sigma$, $\omega$, $\rho$, $\pi$ meson
exchange in the framework of the SU(4) 
chiral invariant effective Lagrangian.
In this work, we use the Lagrangians derived by Ding \cite{ding}
for the interactions $\sigma-\bar{D}D$ and $\omega-\bar{D}D$:

\begin{eqnarray}
\label{model}
{\mathcal{L}}& = 
&(\partial_{\mu}D)(\partial^{\mu}D^{\dagger})-
m_{D}^{2}DD^{\dagger}-\frac{1}{4}F_{\mu\nu}F^{\mu\nu} 
+\frac{1}{2}m_{\omega}^{2}\omega_{\mu}\omega^{\mu}+\frac{1}{2}(\partial_{\mu}\sigma)(\partial^{\mu}
\sigma) 
-\frac{1}{2}m_{\sigma}^{2}\sigma^{2}
\nonumber \\
&&+g_{_{D\bar{D}\sigma}}DD^{\dagger}\sigma
+ig_{_{D\bar{D}\omega}}\omega^{\mu}\lbrack D\partial_{\mu
}D^{\dagger}-(\partial_{\mu}D)D^{\dagger}\rbrack,
\end{eqnarray}
where the field tensor is defined by
$F_{\mu\nu}=\partial_{\mu}\omega_{\nu}-\partial_{\nu}\omega_{\mu}$,
$D$ field is a doublet $(D^{0},D^{+})$,
$m_{D}$, $m_{\sigma}$ and $m_{\omega}$ 
are respectively the masses of the $D$ meson, $\sigma$ meson  
and $\omega$ meson. The coupling constants of this theory
are: $g_{D\bar{D}\sigma}$ and $g_{D\bar{D}\omega}$. 
Applying the Lorentz gauge $\partial_{\lambda}\omega^{\lambda}=0$,
the equations of motion were obtained from Eq.(\ref{model}) are:
\begin{eqnarray}
\label{Em1}
\partial_{\mu}\partial^{\mu}\sigma+m_{\sigma}^{2}\sigma
=g_{_{D\bar{D}\sigma}}DD^{\dagger},
\end{eqnarray}
\begin{eqnarray}
\label{Em2}
\partial_{\mu}\partial^{\mu}\omega^{\rho}+m_{\omega}^{2}\omega^{\rho
}=-ig_{_{D\bar{D}\omega}}[D\partial^{\rho}D^{\dagger}-(\partial^{\rho}D)D^{\dagger
}],
\end{eqnarray}
\begin{eqnarray}
\label{Em3}
\partial^{'}_{\mu}\partial^{'\mu}D
+m_{eff}^{2}D=0,
\end{eqnarray}
where $\partial_{\mu}^{'}=\partial_{\mu} +ig_{_{D\bar{D}\omega}}\omega_{\mu}$ and
$m_{eff}^{2}=m_{D}^{2} -g_{_{D\bar{D}\sigma}}\sigma+
g_{D\bar{D}\omega}^2\omega^{2}$.

The equations Eq.(\ref{Em1}) and Eq.(\ref{Em2}) are
field equations with massive quanta and 
$D$ meson currents as source.
The equation Eq.(\ref{Em3}) is a Klein-Gordon equation 
for the $D$ field with the meson fields $\omega$ and $\sigma$
included in a minimal substitution.

This situation is analogous to Walecka's study of nuclear matter \cite{walecka}, where Walecka's baryon 
field is now represented by $D$ meson field. 

Considering Walecka's mean field theory \cite{walecka}
for an uniform and static system of $D$ mesons 
where the $\sigma$ and $\omega$ meson fields  
can be replaced by classical fields. Thus, the equations 
Eq.(\ref{Em1}), Eq.(\ref{Em2}) and Eq.(\ref{Em3}) reduce to:
%
\begin{eqnarray}
\label{em11}
 \left\langle \sigma\right\rangle =\frac{g_{_{D\bar{D}\sigma}}}{m_{\sigma}
^{2}}\rho_{s},
\end{eqnarray}
\begin{eqnarray}
\label{em22}
 \left\langle \omega^{0}\right\rangle =\frac{g_{_{D\bar{D}\omega}}}
{m_{\omega}^{2}}\rho_{v},
\end{eqnarray}
\begin{eqnarray}
\label{em33}
\partial_{\mu}\partial^{\mu}D+(m_{D}^{2}-g_{_{D\bar{D}\sigma}}\left\langle \sigma\right\rangle
)D+2ig_{_{D\bar{D}\omega}}\left\langle \omega^{0}\right\rangle\partial_{0}D=0,
\end{eqnarray}
where the scalar density is $\rho_{s}=\langle DD^{\dagger}\rangle$ and the vector density is $\rho_{v}=\langle
i[(\partial^{0}D)D^{\dag} -D(\partial^{0}D^{\dag})]\rangle$.
These mean value is done in thermal state and 
integrate these densities in a box with volume $V$ and
divide this result by $V$. 
 
The solution for the $D$ field is given by:
\begin{eqnarray}
\label{field1}
D(\vec{x},t)=e^{-ig_{D{\bar{D}}\omega}\left\langle \omega^{0}\right\rangle t}
{\int}\frac{d^{3}\vec{k}}{2q^{0}(\vec{k})}
\left[a(\vec{k})\frac{e^{-ikx}}{(2\pi)^{3/2}}+
b^{\dagger}(\vec{k})\frac{e^{ikx}}{(2\pi)^{3/2}}\right],
\end{eqnarray}
where 
$kx = q^0x^0-\vec{k}\vec{x}$ and 
$q^{0}(\vec{k})=
\sqrt{
\vec{k}^{2}+m_{eff}^{2}}$.
After the second quantization, the commutation relations 
between the operators are given by:
\begin{eqnarray}
\label{com1}
&&\left[a(\vec{k}),a(\vec{k}^{\prime})\right]
=\left[b(\vec{k}),b(\vec{k}^{\prime})\right]
=\left[a(\vec{k}),b^{\dagger}(\vec{k}^{\prime})\right]
=\left[b(\vec{k}),a(\vec{k}^{\prime})\right]
=0,
\nonumber \\
&&\left[a(\vec{k}^{\prime}),a^{\dagger}(\vec{k})\right]
=\left[b(\vec{k}^{\prime}),b^{\dagger}(\vec{k})\right]
=2q^{^{_{0}}}(\vec{k})\delta^{3}(\vec{k}-\vec{k}^{\prime}).
\end{eqnarray}

The thermodynamic quantities are obtained by
grand potential $\Phi(T,V,\mu)$ at specified chemical potential,
volume and temperature, is defined by:
\begin{eqnarray}
\label{grandepotencial}
\Phi(T,V,\mu)=-\mbox{T} \ln{\Xi},
\end{eqnarray}
where,
\begin{eqnarray}
\label{grandefuncao}
{\Xi}(T,V,\mu)=\mbox{Tr}\left[e^{-(\hat{\mbox{H}}-\mu\hat{\mbox{N}})/\mbox{T}}\right].
\end{eqnarray}
The $\hat{\mbox{H}}$ is the Hamiltonian operator and $\hat{\mbox{N}}$ is the number operator, are given by:
\begin{eqnarray}
\label{op.num} \hat{\mbox{N}}={\int}\frac{d^{3}k}{2q^{0}(\vec{k})}\left\{a^{\dagger}(\vec{k})a(\vec{k})-b^{\dagger}(\vec{k})b(\vec{k})\right\},
\end{eqnarray}
\begin{eqnarray}
\label{op.hamil}
\hat{\mbox{H}}=
\frac{1}{2}\int d^{3}k\left[a^{\dagger}(k)a(k)+b^{\dagger}(k)b(k)\right]
+
g_{D\bar{D}\omega}
\left\langle\omega^{0}\right\rangle
\hat{\mbox{N}}
+
\left(
\frac{m_{\sigma }^{2}\left\langle \sigma \right\rangle^{2}}{2}
-\frac{m_{\omega}^{2}\left\langle \omega ^{0}\right\rangle^{2}}{2}
\right)V,
\end{eqnarray}
where the operators above are the normal-ordered expression.

Inserting the operators $\hat{\mbox{H}}$ and $\hat{\mbox{N}}$
in equation Eq.(\ref{grandefuncao}), 
the grand potential is:
\begin{eqnarray}
\label{g.potential}
\Phi(T,V,\mu)&=&T\frac{V}{2\pi^2}
\int^{\infty}_{0}k^{2}dk 
\left\{
\mbox{ln}\left[1-e^{-\left(
q^{0}(k)-\mu
+
g_{D\bar{D}\omega}
\left\langle\omega^{0}\right\rangle
\right)/T}\right]
+
\left[
1-e^{-\left(q^{0}(k)+\mu -g_{D\bar{D}\omega}
\left\langle\omega^{0}\right\rangle
\right)/T}
\right]          
\right\}\nonumber \\
& &+
\left(\frac{m_{\sigma }^{2}\left\langle \sigma \right\rangle^{2}}{2}
-\frac{m_{\omega}^{2}\left\langle \omega ^{0}\right\rangle^{2}}{2}
\right)V.
\end{eqnarray} 

Finally, we get the mean value of the number operator
\begin{eqnarray}
\label{N.thermal}
\left\langle N\right\rangle=-\frac{\partial\Phi}{\partial\mu},
\end{eqnarray}
matching the $Eq.(\ref{N.thermal})$ with 
the mean value of $Eq.(\ref{op.num})$, we obtain
\begin{eqnarray}
\label{E15}
\left\langle a^{\dag}(\vec{k})a(\vec{k})\right\rangle
=\frac{V}{4{\pi}^3}\frac{q^{0}(k)}
{
e^{\left(q^{0}(k)-\mu+g_{D\bar{D}\omega}\langle\omega^{0}\rangle
\right)/T}
-1},
\end{eqnarray} 
%
\begin{eqnarray}
\label{E16}
{\left\langle b^{\dag}(\vec{k})b(\vec{k})\right\rangle}
=\frac{V}{4{\pi}^3}\frac{q^{0}(k)}{e^{\left(q^{0}(k)+\mu
-g_{D\bar{D}\omega}
\left\langle\omega^{0}\right\rangle
\right)/T}-1}.
\end{eqnarray} 
In $\left\langle N\right\rangle=0$ case, 
we get $\mu=g_{D\bar{D}\omega}\left\langle\omega^{0}\right\rangle$. 
Inserting the mean values given above in normal-ordered Eq.(\ref{em11}), 
we thus obtain the self consistent equation for sigma field:
\begin{eqnarray}
\label{self}
\left\langle \sigma\right\rangle =
\frac{g_{_{_{_{_{DD\sigma}}}}}}{m_{\sigma}^{2}}
\frac{1}{2\pi^{2}}{\int_{0}^{\infty}}
\frac{k^{2}dk}{q^{0}(k)\left[e^{q^{0}(k)/T}-1\right]}.
\end{eqnarray}

Inserting the equations Eq.(\ref{E15}) and Eq.(\ref{E16}) 
in normal ordered Eq.(\ref{em22}), we get the equation for omega field:
\begin{eqnarray}
\label{self2}
\left\langle \omega^{0}\right\rangle =\frac{2g^2_{DD\omega}}{m_{\omega}^{2}}\left\langle \omega^{0}\right\rangle \rho_{s},
\end{eqnarray}
where the solution is given by
\[
\left\langle \omega^{0}\right\rangle =0.
\]
\section{Results}

The parameters used are: $g_{D{\bar{D}}\sigma}=2.85$ GeV \cite{ding}, $m_D=1.87$ GeV and $m_{\sigma}=0.5$ GeV \cite{pdg}.
We use a procedure \cite{Anchishkin:2006nx} to solve the self consistent equation Eq.(\ref{self}) by a method 
to find the roots of the function $F(\langle\sigma\rangle,T)$,
\[
F(\langle\sigma\rangle,T)=\langle\sigma\rangle -LHS(Eq.(\ref{self})).
\]

\begin{figure}[!h]
\begin{center}
\includegraphics[height=7cm]{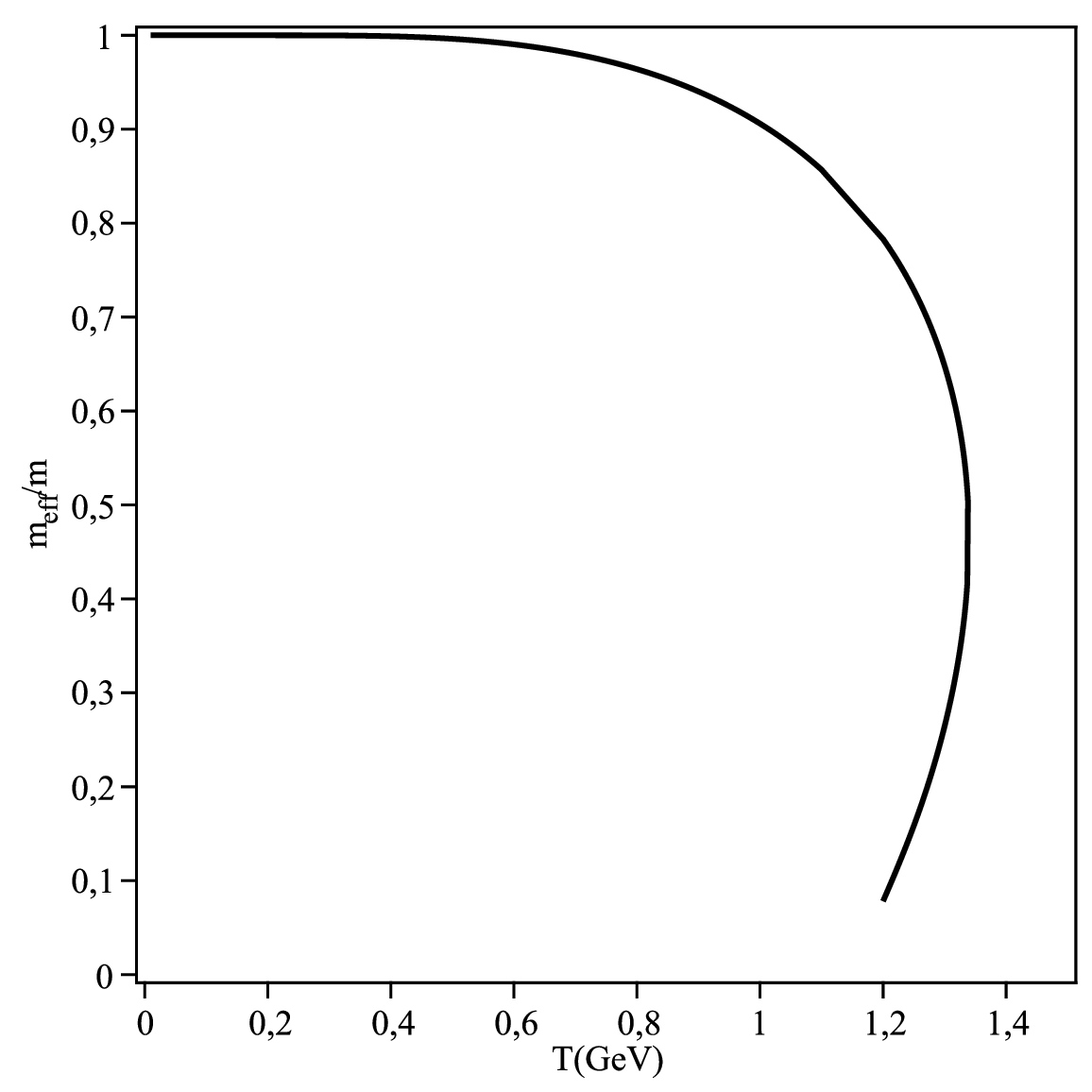}
\caption{\small Effective mass as a function of temperature.}
\label{massef}
\end{center}
\end{figure}

Numerically, we get the 
behavior of sigma field with the temperature. 
Inserting the sigma field in an effective mass equation, 
\[
m_{eff}=\sqrt{m_{D}^{2}-g_{D{\bar{D}}\sigma}
\left\langle \sigma\right\rangle},
\]
we get the mass of the D meson in hadronic 
medium Fig.(\ref{massef}). 
The effective mass reduces with increasing temperature
and has a sudden drop at temperature T $\approx$ 1.2 GeV.
The minimum value of the ratio $m_{eff}/m$ is $0.08$.
This value is much smaller than the value $0.40$
for the pion matter \cite{Anchishkin:2006nx}.
Our result is similar the
nucleon-antinucleon matter \cite{Theis:1984qc}, but
in our result, there is a maximum temperature 
around 1.4 GeV, where there is not any real sigma field. 
In recently paper, Kummar and Mishra \cite{Kumar:2009xc}
predicted a different behavior of the D meson mass  
in nuclear medium. They get the mass grows with increasing temperature.  

\begin{figure}[!h]
\begin{center}
\includegraphics[height=7cm]{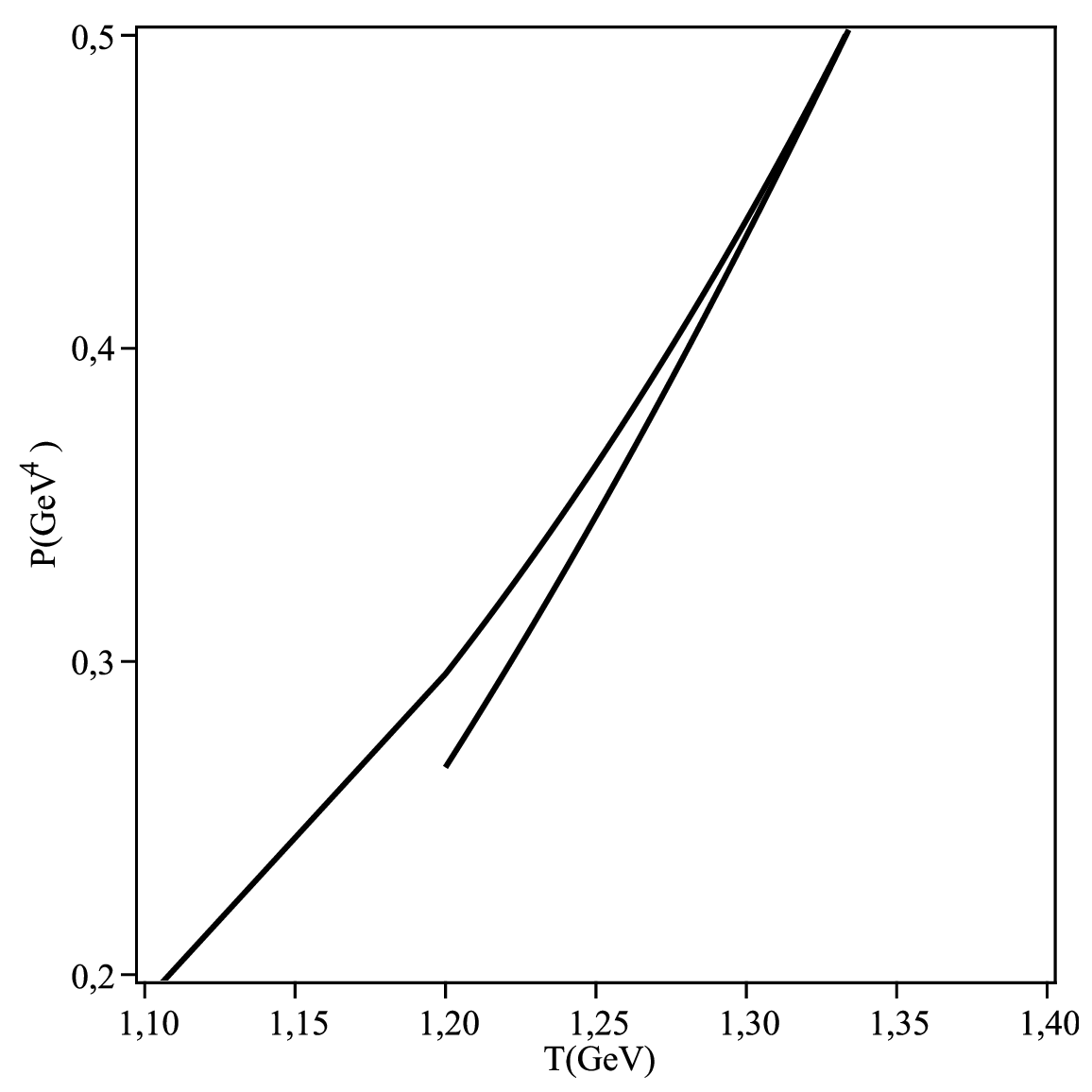}
\caption{\small Pressure as a function of temperature.}
\label{press}
\end{center}
\end{figure}

\begin{figure}[!h]
\begin{center}
\includegraphics[height=7cm]{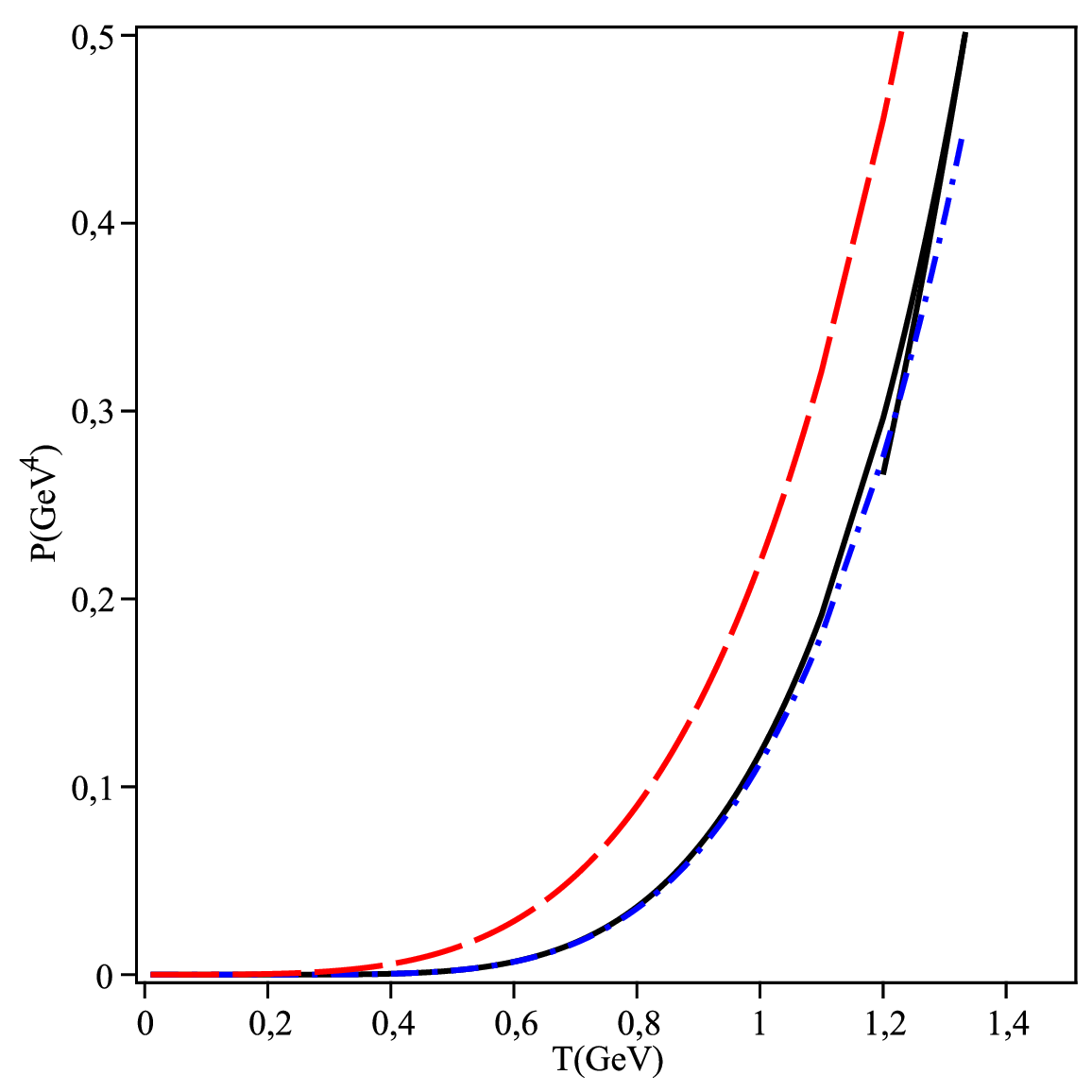}
\caption{\small 
Pressure as a function of temperature.
The solid line shows the
$\sigma \neq 0$ case,
the dot-dashed line shows the
$\sigma=0$ case, the long-dashed line shows the
$m_{D}=\sigma=0$ case.}
\label{press2}
\end{center}
\end{figure}

\begin{figure}[!h]
\begin{center}
\includegraphics[height=7cm]{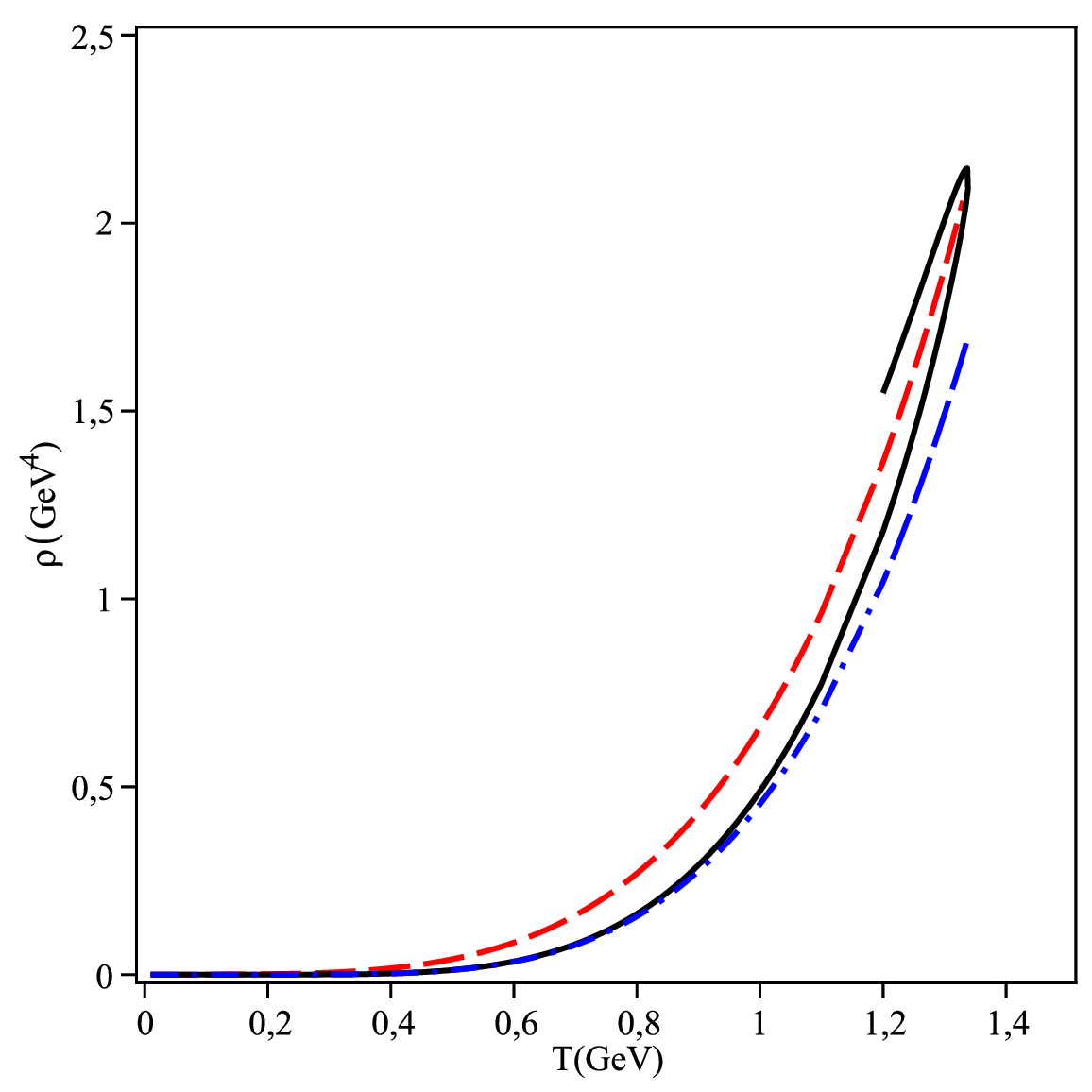}
\caption{\small Energy density as a function of temperature.
The solid line shows the 
$\sigma \neq 0$ case,
the dot-dashed line shows the 
$\sigma=0$ case, the long-dashed line shows the 
$m_{D}=\sigma=0$ case.
}
\label{ener_densi}
\end{center}
\end{figure}

We study the thermodynamics functions in three situations: 
$\sigma \neq 0$, $\sigma=0$ and $m_{D}=\sigma=0$.
The Fig.(\ref{press}) shows the pressure as function
of temperature for the $\sigma \neq 0$ case. It exhibits 
a phase transition around T $\approx$ 1.2 GeV.
The Fig.(\ref{press2}) shows that the other 
two systems have a single phase, 
corresponding to free gas. 
The Fig.(\ref{ener_densi}) shows the behavior 
of the energy density with temperature. 

\begin{figure}[!h]
\begin{center}
\includegraphics[height=7cm]{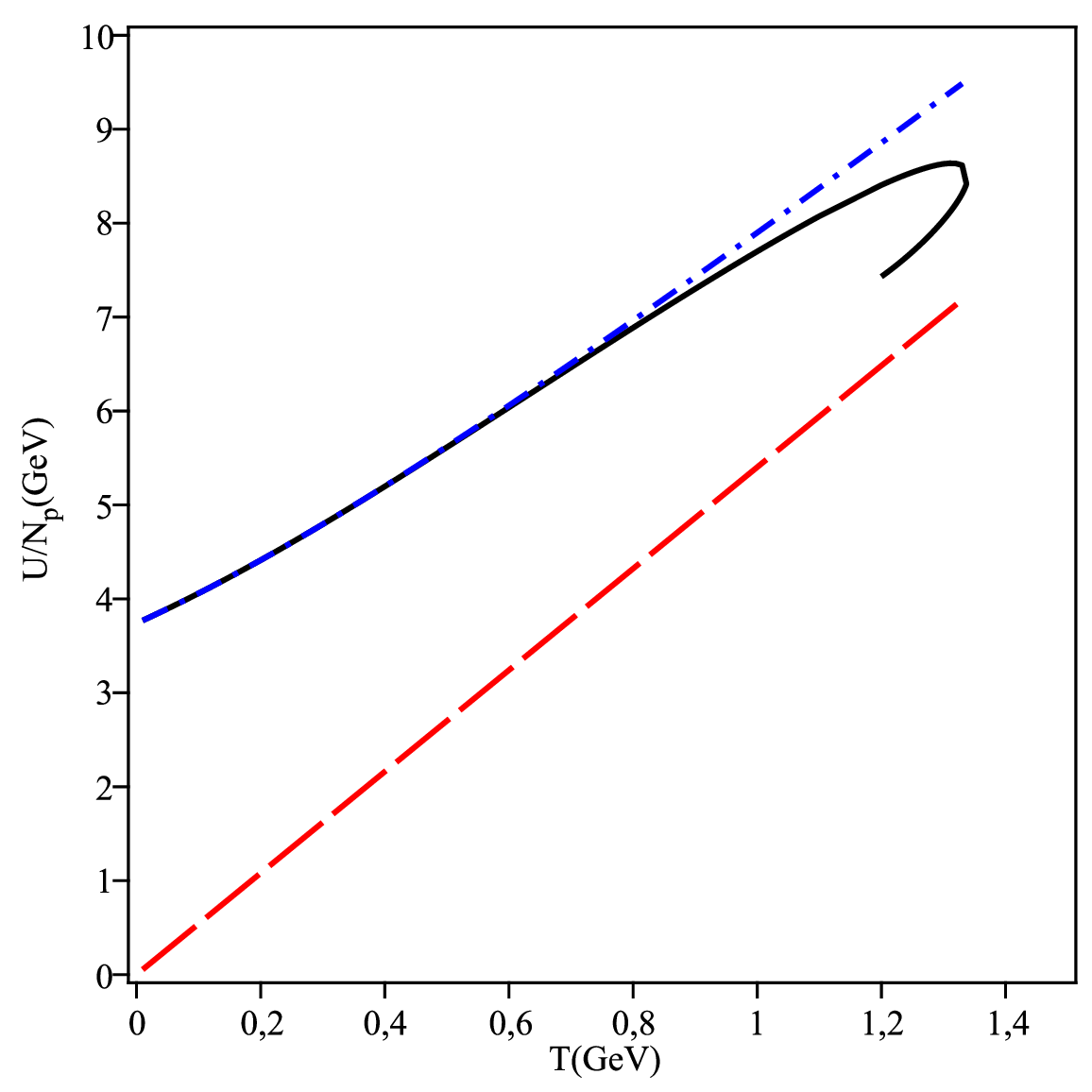}
\caption{\small Energy per pair as a function of temperature.
The solid line shows the 
$\sigma \neq 0$ case,
the dot-dashed line shows the 
$\sigma=0$ case, the long-dashed line shows the 
$m_{D}=\sigma=0$ case.
}
\label{ener_per}
\end{center}
\end{figure}

As our system has the same number of D and anti-D mesons,
we create the quantity energy per pair, where we
divide the total energy per the number of D mesons, $N_{p}$.
The Fig.(\ref{ener_per}) shows the behavior for the
energy per pair with temperature. 
For the $\sigma \neq 0$ case, 
the energy per pair 
has lower values than the $\sigma=0$ case.
This result is in agreement on the fact that 
the sigma field play an important role to create bound states.

These results are analogous to
nucleon-antinucleon matter case \cite{Theis:1984qc}
and support the existence of phase transition 
for the $\sigma \neq 0$ case are 
an indication that this new phase 
is almost free zero mass D mesons Fig.(\ref{massef})
and not a liquid phase. 

{\it Note (2012):
in Ref.\cite{Abreu:2012cc} 
this interpretation has been updated.
}
 

\newpage
\section{Conclusions}

These results show that our system 
is a gas. 
With increasing temperature has a phase 
transition at temperature T $\approx$ 1.2 GeV.
These new state is not a liquid and looks like
almost a free zero mass D mesons gas. 
These results could be interpreted as 
an indication that 
the interaction between $D$ and $\bar{D}$ is not 
so strong to form a liquid phase and 
it becomes difficult to understand
the existence of the $D-\bar{D}$ molecules
or X(3700) in this theory. 

{\it Note (2012):
in Ref.\cite{Abreu:2012cc} 
this interpretation has been updated and
the existence of the $D-\bar{D}$
molecules or X(3700) in this theory
is supported.
}

\section{Acknowledgements}
{We are indebted to Emanuel Cunha and Tomaz Passamani 
for fruitful discussions.
This work has been partially supported by CAPES and CNPq (Brazil).}

\end{document}